\documentclass[
preprint,
 amsmath,
 amssymb,
 aps,
]{revtex4-2}

\usepackage{graphicx}
\usepackage{dcolumn}
\usepackage{bm}
\usepackage{physics}
\usepackage{xcolor}
\usepackage[mathlines]{lineno}

\begin{document}

\title{Stress and Alignment Response to Curved Obstacles in Growing Bacterial Monolayers}

\author{Blake Langeslay$^1$}

\author{Will Fahy$^2$}

\author{Gabriel Juarez$^2$}
 \email{gjuarez@illinois.edu}

\affiliation{$^1$Department of Physics, University of Illinois
Urbana-Champaign, Urbana, Illinois 61801, USA}

\affiliation{$^2$Department of Mechanical Science and Engineering, University of Illinois Urbana-Champaign, Urbana, Illinois 61801, USA}

\date{\today}

\begin{abstract}

Monolayers of growing bacteria, confined within channel geometries, exhibit self-organization into a highly aligned laminar state along the axis of the channel. Although this phenomenon has been observed in experiments and simulations under various boundary conditions, the underlying physical mechanism driving this alignment remains unclear. In this study, we conduct simulations of growing bacteria in 2D channel geometries perturbed by fixed obstacles, either circular or arc-shaped, placed at the channel's center. Our findings reveal that even sizable obstacles cause only short-ranged disruptions to the baseline laminar state. These disruptions arise from a competition between local planar anchoring and bulk laminar alignment. At smaller obstacle sizes, bulk alignment fully dominates, while at larger sizes, planar anchoring induces increasing local disruptions. Furthermore, our analysis indicates that the resulting configurations of the bacterial system display a striking resemblance to the arrangement of hard-rod smectic liquid crystals around circular obstacles. This suggests that modeling hard-rod bacterial monolayers as smectic, rather than nematic, liquid crystals may yield successful outcomes. The insights gained from our study contribute to the expanding body of research on bacterial growth in channels. Our work provides new perspectives on the stability of the laminar state and extends our understanding to encompass more intricate confinement schemes. 

\end{abstract}

\maketitle

\section{\label{sec:level1}Introduction}

Self-organization allows biological active matter at scales ranging from animal flocks to cell skeletons to react and adapt to variable environmental conditions. This self organization can be powered by motility of the individual components of the system \cite{copenhagen, kawaguchi, saw, dunkel, sugi, sanchez}, or it can be powered directly by growth \cite{doostmohammadi, ye, maleki, yaman, liu, hickl1}. A ubiquitous example of a growth-powered active system is a growing colony of non-motile bacteria. Even in the absence of individually directed motion, these colonies can self-organize to adapt to their environments, for example by escaping the monolayer into the third dimension or by simply rearranging to alleviate pressure \cite{maier, farrell, you1, you2, volfson, dellarciprete, boyer, grant, duvernoy, krajnc, basaran}.

Bacterial monolayers have been successfully described as extensile active nematics \cite{dellarciprete, you1, copenhagen}. Active nematic behavior can be reconstructed from a simple model of the bacteria as hard spherocylinders that extend and divide over time. The steric forces between cells produce alignment, and the growth produces extensional motion. However, in some crucial cases, these hard-rod monolayers have also been found to deviate from continuum active nematic behavior. When growing without confinement, bacteria organize into highly-aligned microdomains with alignment discontinuities along their boundaries rather than the point discontinuous topological defects of continuum active nematics \cite{you1, langeslay}. This behavior is characteristic of a close-packed granular material, reflecting the fact that the monolayer is composed of discrete hard elements \cite{donev, kishore}. Additionally, in confinement regimes that induce very high alignment, monolayers have been observed to organize into rows of end-to-end aligned cells \cite{orozco-fuentes, isensee}. Similar behavior can even be seen when looking closely at the microstructure of highly aligned microdomains \cite{you1}. This element of spatial ordering in addition to the orientational ordering of a nematic liquid crystal suggest that these systems may be better characterized as smectic liquid crystals \cite{demus}. Smectic systems can also be constructed from hard rods \cite{frenkel, narayan}, and can produce grain boundaries similar to those observed between bacterial microdomains \cite{monderkamp2, monderkamp}.

One of these highly aligned states can be induced in a growing bacterial monolayer using any confinement scheme in which cells are confined along one axis and unconfined along another \cite{volfson, orozco-fuentes, you2, isensee}. This general class of confinement will hereafter be referred to as a channel geometry. The mechanism for the alignment, in which cells orient along the unconfined axis, is not fully understood. Stress anisotropy was originally proposed to promote cell alignment along the axis of lower stress \cite{you2}. As cells grow confined in one axis and unconfined in another, stress increases along the direction of higher confinement and cells then align along the unconfined direction – that is, parallel to the channel. However, as the system approaches perfect alignment and the system becomes spatially ordered into discrete rows as previously described, this description breaks down \cite{isensee}. Because the active stress in growing-cell systems is oriented entirely along the cells’ long axis, this arrangement prevents any transfer of active stress between rows of cells. This leads to stress decoupling, where the component of stress parallel to the channel is dependent on position (lower near outlets) but the perpendicular component is entirely passive and constant throughout the channel \cite{isensee}. Because of this, in channel centers in very highly aligned systems, the component of stress in the direction of confinement is lower \cite{isensee}. This reversal of stress anisotropy does not result in the rearrangement of cell orientations. The highly channel-aligned state persists even when stress is higher in the direction of alignment.

To better understand this phenomenon, we simulate growing bacteria in channel geometries whose ordering is frustrated by the presence of curved obstacles. In doing so, we draw direct comparisons to the behavior of passive smectic liquid crystals. Like growing bacteria, these can organize into a rich variety of geometric states when confined \cite{wittmann, wittmann2, monderkamp}. These states result from a competition between the constituent particles' drive toward alignment with each other and their drive toward planar anchoring when in contact with a hard boundary \cite{demus, wittmann}, both of which are fully present in a hard-rod bacterial monolayer. In confinement schemes with a circular obstacle, this competition leads to two primary states: a laminar state in which global alignment dominates and planar anchoring is violated (analagous to the highly-aligned state in a bacterial colony), and a Shubnikov state in which planar anchoring is maintained and alignment in the bulk is subject to near-constant bend deformations \cite{wittmann}. 

We find that the highly-aligned laminar state is robust to perturbation. Similarly to passive smectics, bacterial colonies form a laminar state in the presence of circular obstacles. As obstacle size increases this state is disrupted by the increasing dominance of planar anchoring at the obstacle's surface, creating an anchored state where near-obstacle cells align tangent to its surface but more distant cells maintain the laminar state. Additionally, we find that concave obstacles can induce and trap topological defects in the monolayer, potentially leading to novel ways to control the active flow.

\section{Methods}

\begin{figure*}
\centering
    \includegraphics[width=\linewidth]{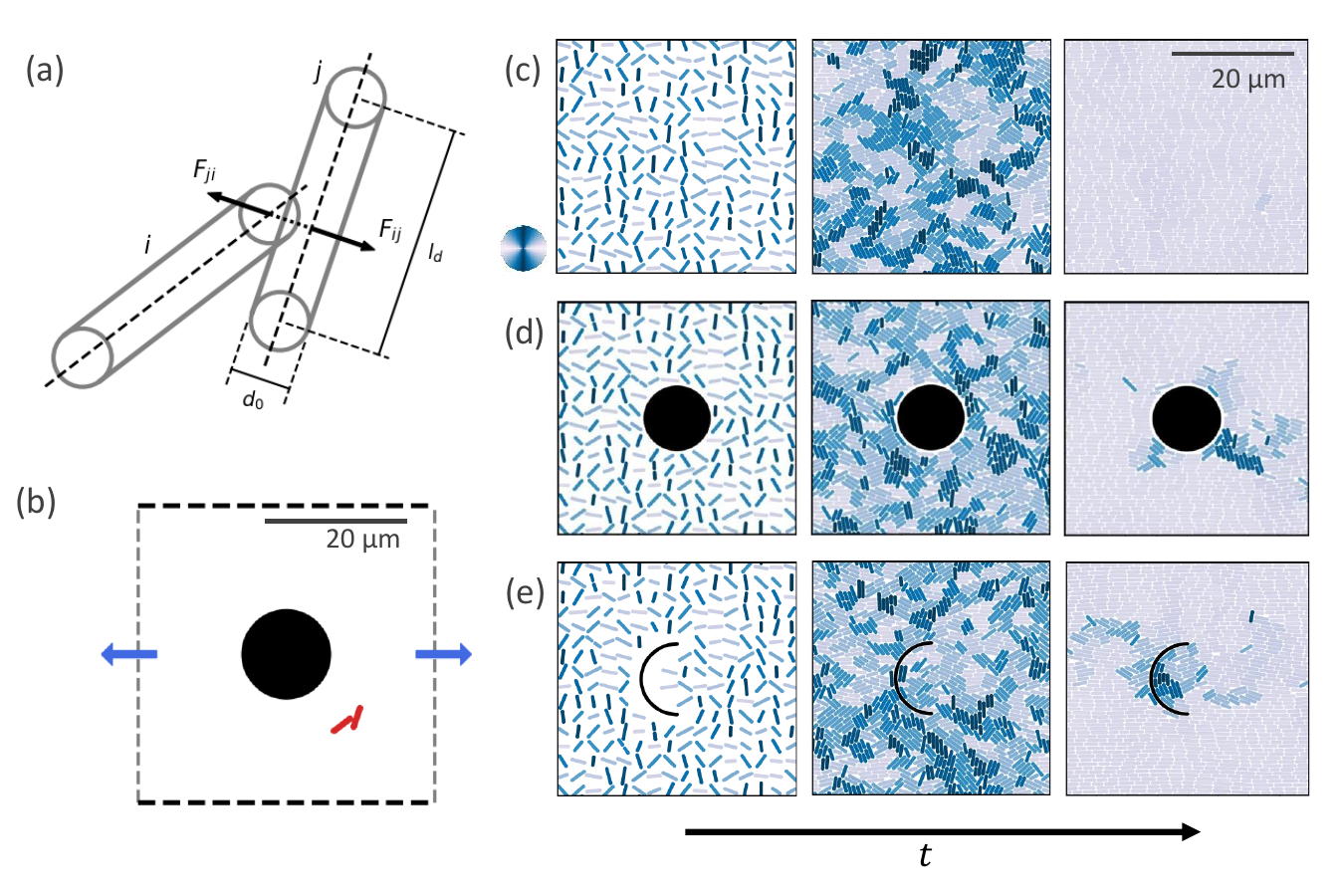}
\caption{
(a) Schematic of cell-cell interaction forces. 
(b) Schematic of the channel confinement geometry with periodic boundaries on top and bottom, outlets on left and right, and a circular obstacle in the center. 
(c-e) Evolution of cell alignment leading up to the steady state for (c) no obstacle, (d) circular obstacle, and (e) arc obstacle simulations. 
Cells are colored by orientation, with light colored cells along the x-axis (horizontal) and dark cells along the y-axis (vertical). 
Cells are shown smaller than their actual size for clarity. 
}
    \label{fig:figone}
\end{figure*}

A molecular dynamics model of cell growth and interaction was used for simulations. This model was based on previous work on similar growing-bacteria systems \cite{orozco-fuentes, you1, isensee, langeslay}. Individual bacteria were modeled as hard, flat spherocylindrical rods constrained to the $xy$-plane. The rods were initialized with length $l$ (defined as the distance between hemispherical endcaps) and set to grow at a uniformly distributed rate between $g_{0}/2$ and $3g_{0}/2$ while maintaining a fixed width of $d_{0}$. Upon reaching the cell-division length of $l_{d}$, a parent cell divided into two child cells, each possessing a unique growth rate from the distribution described previously (Fig. \ref{fig:figone}a). These varying growth rates were used to avoid synchronized cell divisions. A drag-per-unit-length of $\zeta$ was introduced to represent a fluid- or substrate-based drag. Additionally, a noise force $\eta$ was applied to each cell at each time step. 

Apart from noise forces, the translational and rotational motion of the cells were driven exclusively by the physical interactions between contacting cells. This motion was modeled using the following over-damped Newton’s equations:
\begin{equation}
\frac{d\vec{x}}{dt}=\frac{1}{\zeta l}\sum{\vec{F}}
\end{equation}

\begin{equation}
\frac{d\theta}{dt}=\frac{12}{\zeta l^{3}}\sum{\tau}
\end{equation}
where $\vec{F}$ were the forces generated from cell-cell interactions, and $\tau$ were the torques produced about the geometric center of the cell by these same interactions. Intercellular interactions were modeled as Hertzian forces, where overlapping cells generate a repulsive force based on the overlap distance. The force of cell $j$ acting on cell $i$ was defined in the following way: 
\begin{equation}
\vec{F}_{ij}=Yd_{0}^{\frac{1}{2}}h_{ij}^{\frac{3}{2}}\vec{N}_{ij}
\end{equation}
where $Y$ is the cell Young’s Modulus, $h_{ij}$ is the overlap distance between cells, and $\vec{N}_{ij}$ is the vector normal to cell $j$'s surface at the point of contact \cite{you1, hertz}. After each time step, cell positions, lengths and orientations were updated by integrating the equations of motion using the explicit Euler method.

In keeping with the properties of rod-like bacteria and with similar previous simulations \cite{you1, langeslay}, simulation parameters were set to the following values: cell width $d_{0}$ to 0.7 \textmu m, cell division length $l_{d}$ to 2.7 \textmu m, cell growth rate $g_{0}$ to 1 \textmu m/h, Young’s modulus $Y$ to 4 MPa, drag per unit length $\zeta$ to 200 Pa h and maximum noise $\eta$ to 2 $\times$ $10^{-9}$ N. The simulations all used a timestep of 2 $\times$ $10^{-5}$ h, with data recorded at time intervals of 2.5 $\times$ $10^{-2}$ h.

All simulations were run within a square region of side length 40 \textmu m. The upper and lower boundaries were periodic, while the left and right boundaries were outlets. Cells passing an outlet boundary were removed from the simulation. A schematic of the simulation setup is shown in Fig. \ref{fig:figone}b. The system was initialized with 256 cells spaced evenly across the area of the square in a 16 by 16 grid with random initial orientations and lengths. 

Obstacles were modeled as fixed objects within the simulation area, exerting a Hertzian force on overlapping cells identical to the cell-cell interaction force. The obstacle simulations were divided into two subsets: simulations with a fixed circular obstacle centered at the origin, and simulations with a semicircular arc-shaped obstacle with its radial center fixed to the origin. For the circular obstacle simulations, circles of radius $R$ = 2, 4, 6 and 8 \textmu m were tested. The arc obstacles had width 0.7 \textmu m (the same as a cell) with semicircular endcaps. An arc’s radius was defined as the distance from its radial centerpoint to its centerline. For the arc obstacle simulations, arcs of radius of 4, 6, and 9 \textmu m were tested. To avoid non-physical interactions, cells that would have initialized within the obstacle boundary were removed before the simulation began.

\section{Results}

In each simulation, the initially scattered cells grow and divide to eventually cover the entire surface of the 2D channel (Fig. \ref{fig:figone}c-e, supplemental videos SV1-3 \cite{supp}). As the cells populate the surface, they exert contact forces on neighboring cells, generating greater alignment among neighbors and resulting in the formation of a locally nematically ordered liquid crystal \cite{dellarciprete, you1}. This creates a period during which cells have fully crowded the substrate but are still poorly aligned on the global scale (Fig. \ref{fig:figone}c-e, middle column). 

Once the monolayer has fully covered the surface, the continued growth of the cell monolayer combined with the periodic confinement in the $y$ direction generates an extensional flow, pushing cells from the channel center ($x=0$) towards the outlets [obstacle supplemental video(s)]. At long times, all simulation geometries drive global alignment of the monolayer in the horizontal direction producing a highly aligned phase as seen in previous channel-like experiments and simulations (Fig. \ref{fig:figone}c-e, right column) \cite{volfson, orozco-fuentes, isensee}. Unless otherwise stated, all following results are averaged over time in the highly aligned phase ($t>16$ h). We will first investigate the alignment and stress distribution within the highly aligned monolayer produced by circular obstacles, followed by the alignment and velocity profiles produced by arc obstacles.

\subsection{Circular Obstacles}

At high alignment, the system forms discrete rows of parallel, end-to-end cells, visible in simulation images (Fig. \ref{fig:figone}c-e, right column). These rows form anywhere the system is highly aligned, spanning the entire system in unobstructed or small-obstacle simulations but only forming farther away from larger obstacles (Fig. \ref{fig:figone}d, right column). As seen in previous channel-geometry research \cite{isensee}, these discrete rows add an element of spatial ordering to the nematic order common in bacterial monolayers.

To track the progression of system-wide behavior in the presence of circular obstacles, the local nematic orientational order was calculated. The local order $S_{i}$ of cell $i$ is given by the following equation:
\begin{equation}
S_{i}=\frac{1}{N}\sum_{i}{2}\cos^{2}(\theta_{i}-\theta_{j})-1
\end{equation}
where $\theta_{j}$ is the orientation of each cell $j$ whose center lies within a search radius of cell $i$ and $N$ is the number of cells within the search radius. The parameter is structured such that $S\rightarrow 1$ denotes perfect alignment, $S\rightarrow 0$ denotes no ordering, and  $S\rightarrow -1$ denotes anti-aligned regions. For this analysis, the search radius was set equal to $l_{d}$, the cell division length.

\begin{figure*}
\centering
    \includegraphics[width=\linewidth]{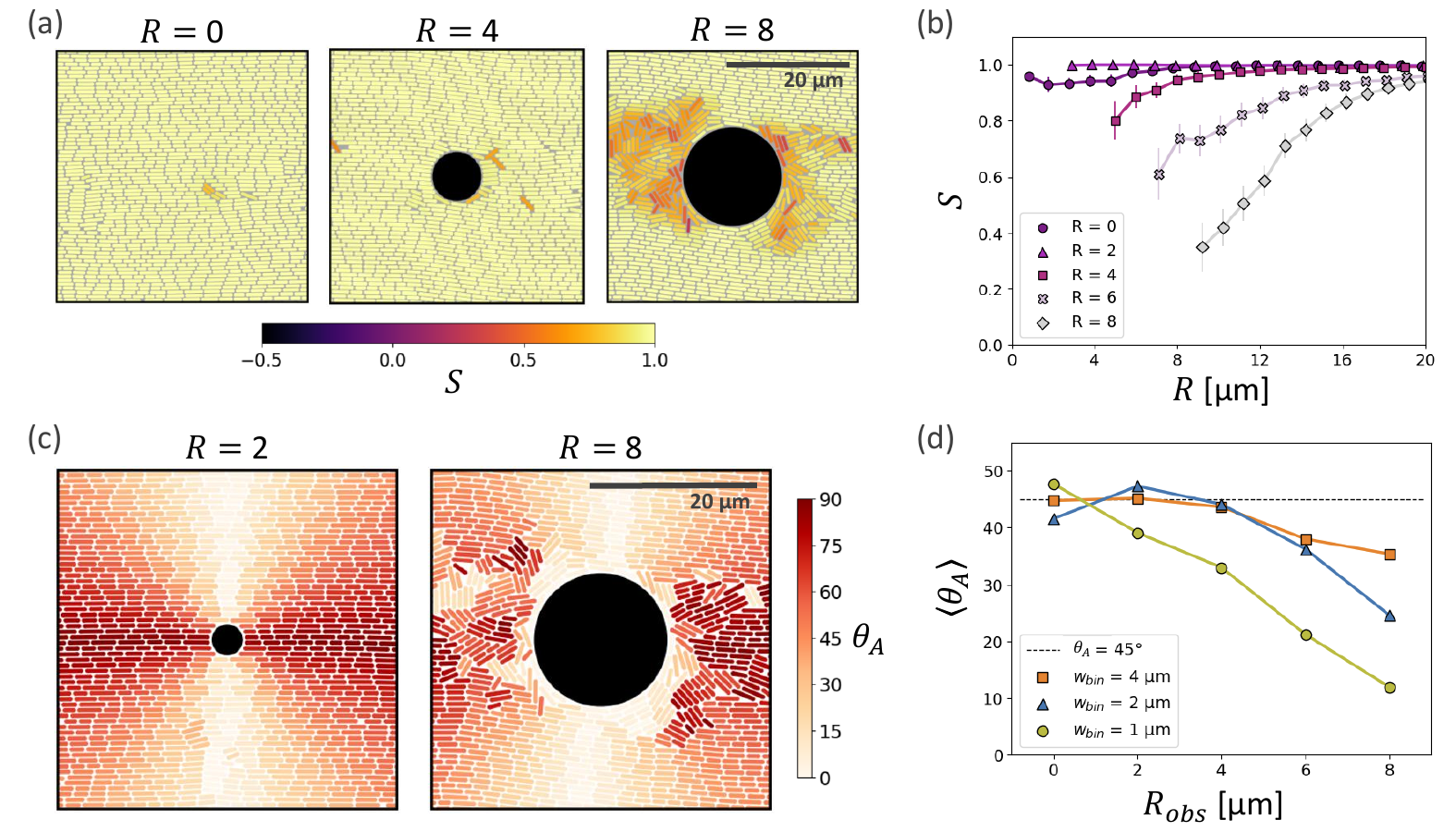}
\caption{
(a) Sample images of cells colored by orientational order (S) around circular obstacles. (b) Average orientational order in the steady state as a function of distance $R$ from the origin. (c) Sample images of simulations with cells colored by anchoring angle. Light colored cells are tangent to the obstacle edge, while dark cells are perpendicular. (d) Average anchoring angle in cells within a distance $w_{bin}$ of the obstacle, as a function of obstacle radius $R_{ob}$.
}
    \label{fig:figtwo}
\end{figure*}

At the beginning of the disordered close-packed phase for all obstacle sizes, local order is homogeneously distributed and low ($S \simeq 0.2$) reflecting the system’s disordered state. As cells continue to grow, local order increases and the system eventually reaches the highly-aligned state (Fig. 2a). Differences in the high-alignment state can be seen between the obstacle sizes. No obstacle ($R=0$) produces globally near-perfect order resulting in a laminar state. Smaller obstacles ($R=4$ \textmu m) produce a near-laminar state interrupted by a few scattered disordered cells near the obstacle’s edge. Larger obstacles ($R=8$) \textmu m produce large regions of disorder around the obstacle reaching to the outlets, with near-perfect alignment appearing only farther above and below the obstacle.

To investigate the spatial distribution of order in the monolayer, average order was calculated in a series of annular bins of width 1 \textmu m around the obstacle (Fig. \ref{fig:figtwo}b). Increasing obstacle size both decreases the local order near the obstacle and increases the range of low-order regions from the obstacle’s surface. For no obstacle and $R=2$, the order remains constant close to 1 for the entire simulation area. For $R=4$, there is a short-range decrease in order within a few cell lengths of the obstacle’s boundary. For $R=6$ and $R=8$, the order near the obstacle is much lower, and low-order regions persist many cell lengths away.

These differences can be better understood by investigating how well cells adjacent near the obstacle align tangent to its curved surface due to planar anchoring. This was quantified by the anchoring angle $\theta_{A}$ between the cell and the tangent line, with $\theta_{A}=0^{\circ}$ denoting perfect planar anchoring and $\theta_{A}=45^{\circ}$ denoting uncorrelated alignment (no enforced anchoring). For the smallest obstacles ($R=2$), cells’ global horizontal alignment is not disrupted at all, resulting in discrete rows extending to the right and left of the circle without any response to its curvature (Fig. \ref{fig:figtwo}c, $R=2$). For the largest obstacles, in contrast, nearly all adjacent cells exhibit planar anchoring (Fig. \ref{fig:figtwo}c, $R=8$).

To quantify the effect of obstacle size on alignment, the average anchoring angle $\langle\theta_{A}\rangle$ was calculated in a series of annular bins of increasing width $w_{bin}$ around the obstacle's edge. In the closest cells ($w_{bin}=1$ \textmu m), $\langle\theta_{A}\rangle$ approaches zero (planar anchoring) roughly linearly with increasing obstacle radius (Fig. \ref{fig:figtwo}d). For cells farther away from the obstacle ($w_{bin}=2, 3$ \textmu m), obstacles of radius 4 \textmu m and lower do not affect $\langle\theta_{A}\rangle$. Increasing the obstacle radius past this point results in increasing tangent alignment in these more distant cells as well as in the nearest cells.

To quantify forces within the monolayer, the stress tensor of each cell was calculated \cite{you1}. The stress $\boldsymbol{\sigma}_{i}$ on cell $i$ is given by the following equation:
\begin{equation}
\boldsymbol{\sigma}_{i}=\sum_{i}{r_{ij}F_{ij}}
\end{equation}
where $r_{ij}$ is the vector from the center of cell $i$ to the point of contact on cell $j$, and $F_{ij}$ is the force of cell $j$ on cell $i$. The components $\sigma_{xx}$ and $\sigma_{yy}$ represent the stresses along and across the channel, respectively. These components were then averaged in a $12 \times 12$ grid of square bins. For each simulation, the stresses were normalized by $\sigma_{m}$, the highest bin-averaged value of $\sigma_{xx}$ in that simulation.

\begin{figure*}
\centering
    \includegraphics[width=\linewidth]{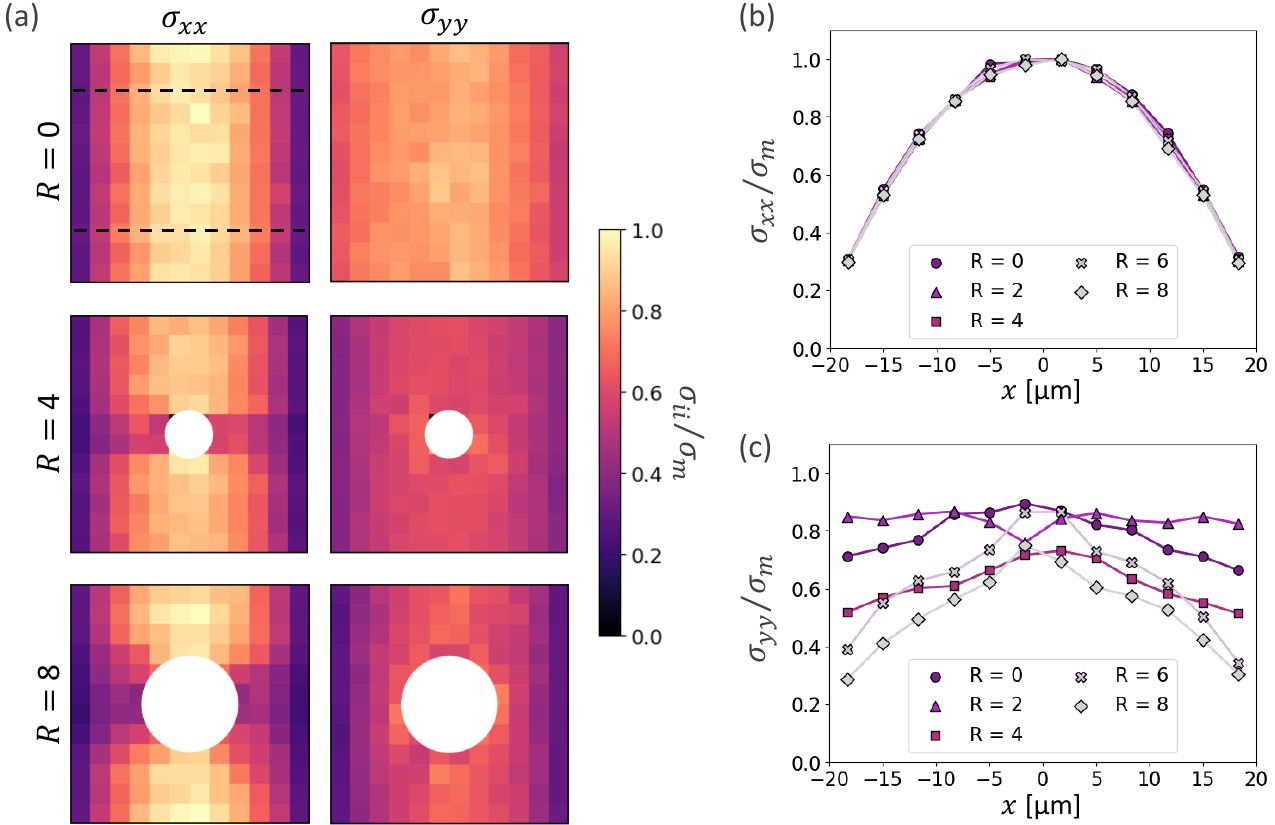}
\caption{
(a) Heatmaps showing steady-state stress components $\sigma_{xx}$ and $\sigma_{yy}$ around circular obstacles, normalized by the maximum $\sigma_{xx}$ for the respective simulation. Dotted lines in upper left show bounds for the region in which stress profiles are calculated.
(b) Profile of $\sigma_{xx}$ along the $x$-axis for $\abs{y}>12$ \textmu m (dotted lines in (a)) in each simulation. 
(c) Profile of $\sigma_{yy}$ along the $x$-axis for $\abs{y}>12$ \textmu m in each simulation.
}
    \label{fig:figthree}
\end{figure*}

In all cases, both stress components are highest near $x=0$ and lowest near the channel outlets, with this variation always greater in $\sigma_{xx}$ than in $\sigma_{yy}$ (Fig. \ref{fig:figthree}a). Above and below the obstacle, $\sigma_{xx}$ maintains a similar distribution independent of obstacle size, but to the sides of an obstacle $\sigma_{xx}$ is significantly lower, consistent with the obstacle partially blocking extensional flow toward the outlets in the $x$ direction (Fig. \ref{fig:figthree}a, left column). In contrast, $\sigma_{yy}$ does not depend on position relative to the obstacle, but is globally affected by the obstacle’s presence and size (Fig. \ref{fig:figthree}a, right column). With no obstacle $\sigma_{yy}$ barely varies in the $x$ dimension, only slightly decreasing near the outlets. However, with larger obstacles the difference in stress between the center and outlets increases.

To quantify the global changes in stress distribution induced by the obstacle, the profiles of average stress along the channel away from the obstacle ($\abs{y}>12$ \textmu m) were used (Fig. \ref{fig:figthree}b and \ref{fig:figthree}c). The distribution of $\sigma_{xx}$ along the channel’s length is always parabolic (Fig. \ref{fig:figthree}b) consistent with previous work in both growing rod and continuum nematic simulations \cite{you2, isensee}. For obstacles with $R>2$, $\sigma_{yy}$ has a similar profile with a lower maximum stress magnitude (Fig. \ref{fig:figthree}c). However, with $R=2$ or no obstacle, $\sigma_{yy}$ varies very little along the channel’s length. 

For no obstacle and $R=2$, the lack of variation in $\sigma_{yy}$ with $x$-position while $\sigma_{xx}$ retains a parabolic profile implies that the stress components have become decoupled \cite{isensee}. Larger obstacles cause $\sigma_{yy}$ to be more $x$-dependenct, implying coupling of $x$- and $y$-stresses. This agrees with previous work that showed similar stress decoupling to be a result of high alignment and particularly of system-wide organization into discrete rows \cite{isensee}.

\subsection{Arc Obstacles}

\begin{figure*}
\centering
    \includegraphics[width=3.5 in]{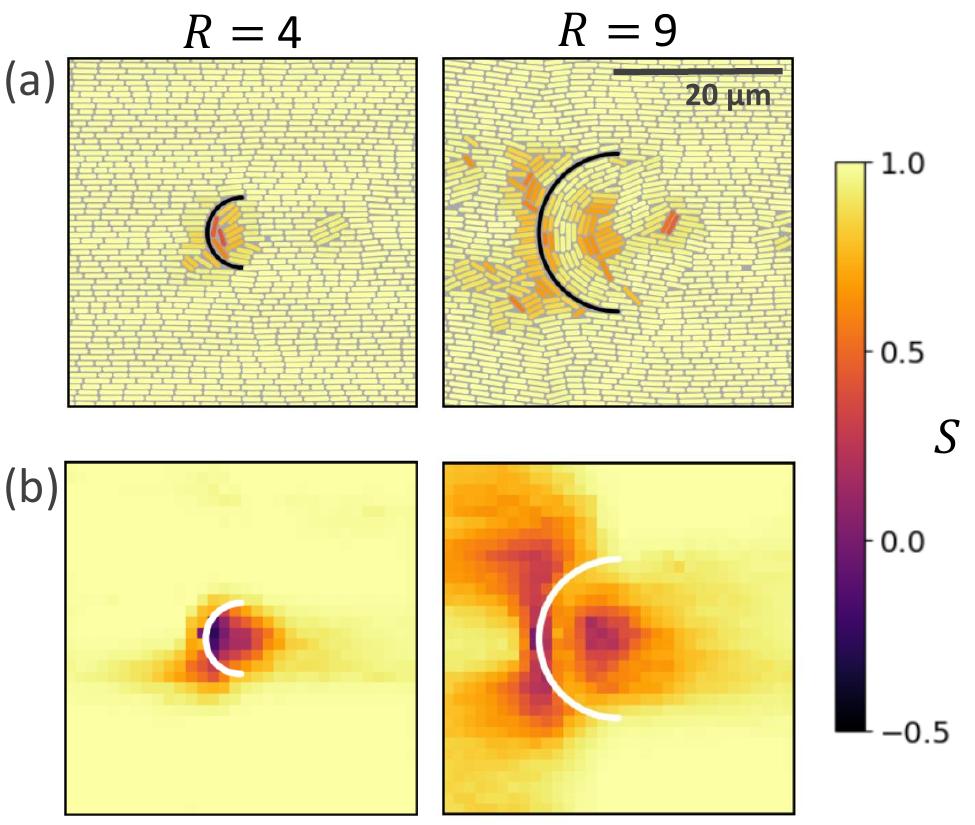}
\caption{
(a) Sample images of steady-state cell alignment (colored by local order $S$) of cells around arc obstacles. 
(b) Heatmaps of local order around arc obstacles. 
}
    \label{fig:figfour}
\end{figure*}

To investigate the effect of concave obstacle curvature we now turn to arc-shaped obstacles. Snapshots of the resulting system configurations around two differently sized arcs are shown in Fig. \ref{fig:figfour}a. Cells adjacent to the interior (right) of arc-shaped obstacles align tangent to the arc's curve. To the left of the arc, cell alignment is similar to that near a circular obstacle, and elsewhere the monolayer maintains its horizontal global alignment with the same discrete-row arrangement described earlier. The combination of arc-aligned interior alignment and horizontal exterior alignment creates a structure reminiscent of a comet-like $+1/2$ charge nematic topological defect \cite{giomi}.

To better understand this defect-like structure, orientational order was averaged in a grid of square bins throughout the highly aligned phase. Order is low within the arc’s curvature, matching the low order in the core of a topological defect (Fig. \ref{fig:figfour}b). For smaller arcs ($R=4$ and $R=6$), the lowest order occurs at the leftmost interior point, with order increasing to the right. In contrast, for the largest arcs ($R=9$), there is a band of higher order close to the arc, with the low-order region localized several cell lengths farther right. This band of high order corresponds to a multiple-cell-wide region of alignment to the larger arc, in contrast to smaller arcs where only adjacent cells align and no longer-range order is produced. Low-order regions also appear to the left of the arc, again matching the response to circular obstacles. As in the circular case, smaller obstacles produce localized disorder near the arc while larger obstacles produce more long-ranged disorder.

\begin{figure*}
\centering
    \includegraphics[width=\linewidth]{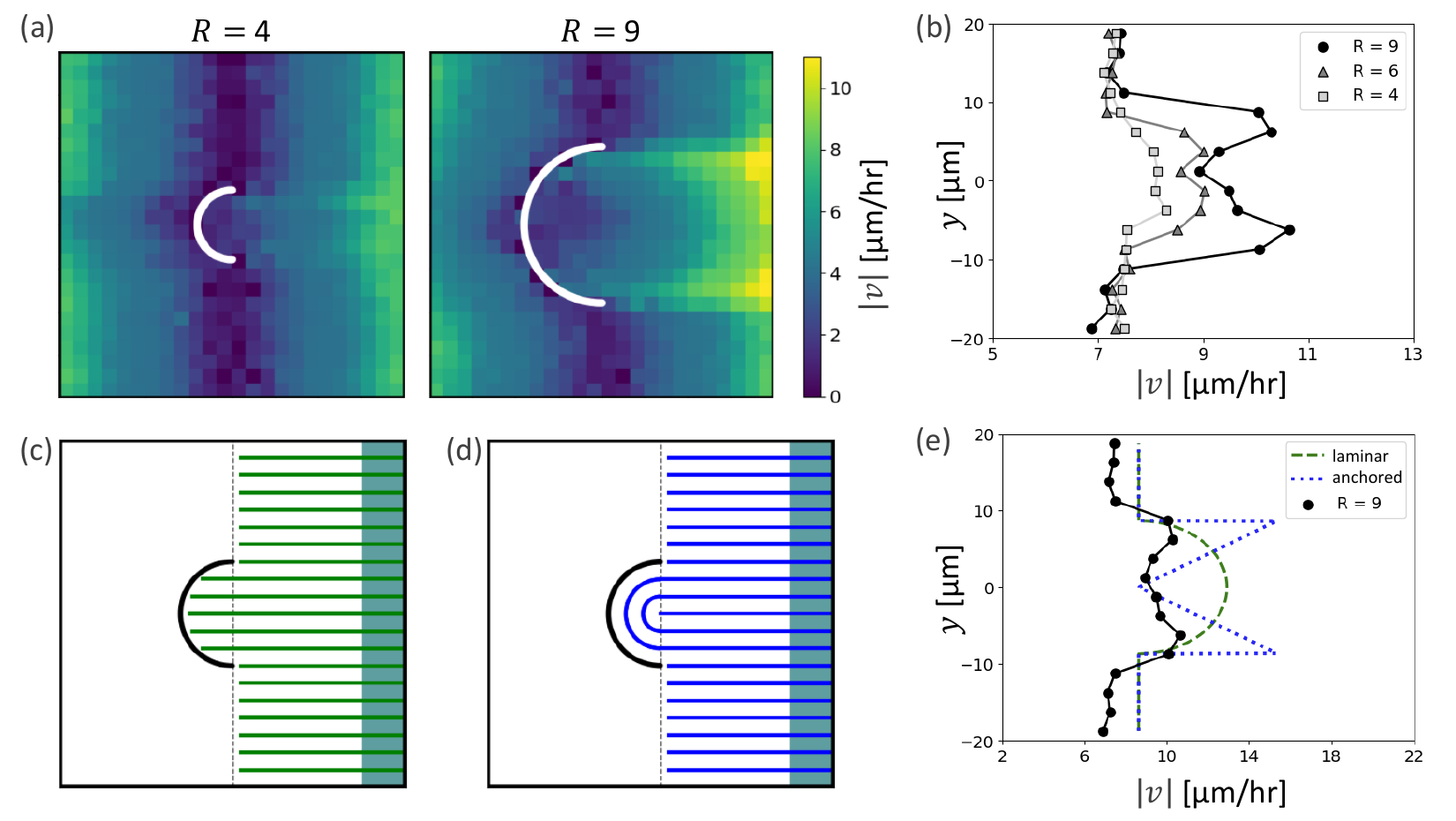}
\caption{
(a) Heat maps of cell velocity magnitude around differently sized arc obstacles. 
(b) Plot of cell velocity magnitude by y position near the right outlet ($x>15.00$ \textmu m). 
(c-d) Visualization of (c) laminar and (d) anchored arrangement of cell rows. 
(d) Comparison plot between proposed laminar and anchored velocity profiles and actual profile for $R=9$ \textmu m simulation.
}
    \label{fig:figfive}
\end{figure*}

Arc-shaped obstacles qualitatively change the motion of cells in the colony. In simulations without obstacles, cells at the center line ($x=0$) have zero average velocity due to the symmetrical nature of the extensional flow (supplemental video SV1 \cite{supp}). Cells away from $x=0$ move toward the closest channel outlet with velocity magnitude increasing with proximity to the outlet, consistent with an unobstructed active extensional flow. The same behavior is maintained away from obstacles when they are present (supplemental video SV2 \cite{supp}). This can be seen in heatmaps of average velocity magnitude for arc obstacle simulations (Fig. \ref{fig:figfive}a). Above and below the arcs, the velocity magnitude is zero at the $x=0$ line and increases with distance from that center line as before, to a maximum of about 7 \textmu m/hr at the outlets.

On the open side of the arcs (in the $+x$ direction), they produce jets of higher-velocity cells exiting the channel (Fig. \ref{fig:figfive}a, supplemental video SV3 \cite{supp}). The velocity profiles of these jets near the outlets ($x>15$ \textmu m) show that their speed increases with increasing arc size (Fig. \ref{fig:figfive}b). The smallest arcs ($R=4$) produce a single-peaked jet velocity profile with a maximum velocity of 8 /textmu m/hr). In contrast, the largest arcs ($R=9$) produce a double-peaked velocity profile with maxima of $v \simeq 11$ /textmu m/hr localized at $y=\pm 9$ \textmu m, corresponding to the arc's edges.

The underlying structure of these jets can be understood with a simple model based on the fact that the system aligns in parallel rows of end-to-end cells. Because each individual cell is growing, when one end of the row is fixed, the other end has a velocity proportional to the number of cells in the row, $v=g_{0}N$. When both ends of the row reach outlets, its center can be treated as fixed as it receives equal force from both sides, and the end cells instead have velocity $v=g_{0}N/2$. Two simple models based on this velocity calculation can explain the observed speed increase. In the laminar model (Fig. \ref{fig:figfive}c), cells interior to the arc retain perfect channel-parallel bulk alignment, resulting in rows with one fixed end of length $L_{i}=L/2+\delta x_{i}$ where $L$ is the channel length and $\delta x_{i}$ is the arc’s depth at the height of the chosen row. In the anchored model (Fig. \ref{fig:figfive}d), interior cells instead align perfectly tangent to the curve of the arc, resulting in curved rows with two free ends of total length $L_{i}=L+\pi \abs{y_{i}}$, where $y_{i}$ is the vertical position of the row. 

By setting $N=L_{i}/<l>$, where $<l>$ is the average cell length including endcaps, we can predict the velocity profiles resulting from the two models. The laminar model predicts a higher velocity exiting the arc’s center, whereas the anchored model predicts a higher velocity exiting its upper and lower edges (Fig. \ref{fig:figfive}e). The laminar model better describes the smallest arc’s behavior, whereas the anchored model fits better with increasing arc size.

\section{Discussion}

Our results show that channel-parallel laminar alignment is robust to perturbation. Even with circular obstacles of diameter nearly half the channel width, cells above and below the obstacle retain a high degree of horizontal alignment and form parallel rows of end-to-end cells. While previous work has simulated global alignment in channels with a variety of different boundary conditions \cite{you2, isensee}, this is the first evidence that it can arise with geometric constraints actively competing with the channel confinement. 

Additionally, we recapitulate the finding that pressure anisotropy can reverse without loss of channel-parallel alignment \cite{isensee}. With the addition of circular obstacles of any size, cells in the channel’s center still experience higher pressure parallel to the channel than perpendicular, and maintain high average alignment parallel to the channel regardless. This happens even in the presence of large obstacles that significantly disrupt order in their vicinity. In other words, the highly-aligned state with higher stress parallel to the channel remains stable in the presence of physical perturbation away from perfect alignment and is not simply an unstable equilibrium state achievable only in systems with a global laminar state.


Cells near a circular obstacle experience competing alignment pressures between bulk laminar alignment and planar anchoring to the obstacle. Bulk alignment dominates at low obstacle sizes, while planar anchoring has increasing effect with increasing obstacle size. For the smallest circular obstacles ($R=2$), the system is fully laminar with no deviation from the no-obstacle case. For larger obstacles ($R=4$), a few cells in contact with the obstacle are captured by planar anchoring, but without meaningfully disrupting the global laminar order. In the $R=6$ and $R=8$ cases near-obstacle cells consistently display planar anchoring, and this more significant break from the global laminar state produces larger disordered regions to the right and left of the obstacle. This qualitative variation in behavior occurs in response to obstacle size variation on the scale of only a few cell lengths.

The case of the smallest circular obstacle is particularly interesting. Images of cell alignment near circular obstacles with $R=2$ show that system-spanning horizontal rows of end-to-end cells occur adjacent to and directly above and below the obstacle. Between these rows, a fixed number of cell rows (6) occur to the obstacle’s right and left, ending at its surface. This configuration leads to a global horizontal alignment that is not just similar to the no-obstacle case but slightly higher (Fig. \ref{fig:figtwo}b). Similarly, this configuration exhibits a higher degree of stress decoupling than the no-obstacle case (Fig. \ref{fig:figthree}c). We suggest that the small obstacle has a stabilizing effect on the highly aligned discrete rows, as it locks the rows above and below it in place and prevents any vertical fluctuations in those cells’ motion. This would explain its anomalous alignment and pressure behavior. It also implies an interesting avenue for further exploration in growing-cell systems whose parameters are conducive to buckling, and where the inclusion of small fixed obstacles might suppress that buckling.

The increase in planar anchoring with obstacle size is likely due to a combination of two factors. First, larger circular obstacles have lower surface curvature, and therefore a planar anchored cell tangent to their surface has less rotational freedom and is more stable. Second, if one considers a laminar configuration around a larger circle, the row just below (above) the top (bottom) locked row would be required to contact the circle with a cell endpoint at a small angle to the obstacle’s surface. This would impart a torque onto that cell that must be countered by the horizontally-aligned collective to prevent rotation, with said torque increasing with circle size. Our results suggest that the critical obstacle size where these effects dominate and perfect laminar alignment near the obstacle no longer occurs is on the scale of 1 to 2 cell lengths.

Concave curved obstacles differ from convex curved obstacles in that straight cells adjacent and tangent to the concave surface are always stable with two points of contact, whereas cells tangent to convex obstacles have only one point of contact and are unstable. Because of this, cells in contact with the interior of an arc are always planar anchored, even in the smallest cases (Fig. \ref{fig:figfour}a). In combination with the global horizontal alignment (which remains undisturbed by the arc’s presence elsewhere), this forces a structure similar to a +1/2 nematic topological defect. 

With channel-parallel alignment far from the arc obstacle, the total topological charge is constrained to be zero. The natural way to achieve this is with a -1/2 defect to the left of the arc, resulting in a pair of +1/2 and -1/2 defects which here are stabilized and prevented from mutually annihilating by the presence of the arc. In practice, only the largest arcs ($R=9$) show two distinct low-order defect cores separated by a region of higher order parallel to the arc (Fig. \ref{fig:figfour}b). Smaller arcs instead produce a single disordered core, indicating that the anchored behavior of adjacent cells does not extend to further curved rows of curved cells. We interpret this to mean that the $R=9$ case produces a well-defined +1/2 defect, while smaller arcs produce more complicated granular disorder in which many cells remain horizontally aligned. This is further supported by the fact that velocity profiles produced by smaller arcs are better modeled by laminar alignment while velocity profiles produced by larger arcs are better modeled by an anchored state. We can describe this in the same simple terms as the circular obstacle case: laminar bulk alignment dominates in the small obstacle (high curvature) case, while planar anchoring becomes more important around larger (less curved) obstacles.

Less specific to the particular smectic nature of the growing-rod model, it is intriguing that fixing a +1/2 defect produces backflow jets. As +1/2 defects in extensional active nematics are motile \cite{giomi}, it seems natural that preventing their movement would consistently result in an equal and opposite active force repelling cells behind them. While we have explained this behavior in terms of the specific configurations of growing bacteria systems, it would be interesting to see a more general investigation of concave obstacles in continuum active nematics, either in theory or in experimental systems such as kinesin-driven microtubules. This is an especially exciting idea in the context of other work on controlling the flow of active nematics \cite{norton, thijssen}.

Our simulations have extended previous work on growing bacteria in channel geometries to show that global laminar alignment and reversal of stress anisotropy are robust under perturbation of confinement geometry. Additionally, we have shown that increasing obstacle size (decreasing obstacle curvature) induces a transition from laminar alignment to states where planar anchoring disrupts said laminar alignment. This behavior has striking similarity to the behavior of passive smectics around obstacles \cite{wittmann, wittmann2}, further supporting the proposition that a smectic description of a growing bacterial monolayer has value. These simulations are a step toward extending the channel geometry understanding of cell alignment to more complicated environments, whether modeling cell behavior on naturally occurring rough substrates or designing custom surfaces to control cell behavior.

\section{Conflicts of Interest}
There are no conflicts of interest to declare.

\begin{acknowledgments}

This work used Bridges-2 at at Pittsburgh Supercomputing Center through allocation phy210132 from the Advanced Cyberinfrastructure Coordination Ecosystem: Services \& Support (ACCESS) program, which is supported by National Science Foundation grants \#2138259, \#2138286, \#2138307, \#2137603, and \#2138296 \cite{bridges2, access}.

\end{acknowledgments}

\bibliography{apssamp}

\end{document}